\documentclass[letter,twocolumn]{jpsj3}
\usepackage[dvipdfmx]{graphicx}

\makeatletter 
\newcommand\newblock{\hskip .11em\@plus.33em\@minus.07em}
\makeatother

\usepackage{amsmath, amssymb}
\usepackage{bm}
\usepackage{mathrsfs}
\usepackage{physics}
\usepackage[normalem]{ulem}

\usepackage{color}

\title{Ferroaxial Electronic Response to Local Lattice Rotation}
\author{Ken Uchino, Yuuki Ogawa, and Satoru Hayami}
\inst{Graduate School of Science, Hokkaido University, Sapporo 060-0810, Japan}

\abst{
We theoretically investigate electronic responses to lattice deformation in a ferroaxial system using a tight-binding model on a two-dimensional square lattice and linear response theory. 
While conventional elastic responses are commonly discussed in terms of symmetric strain, the displacement gradient also contains an antisymmetric component describing a local lattice rotation. 
Focusing on this rotational degree of freedom, we treat the symmetric strain and antisymmetric local rotation as independent perturbations and incorporate their effects microscopically through deformation-induced modulations of the electronic hopping amplitudes.
We find that the local rotation induces characteristic diagonal electric quadrupoles that are distinct from those generated by the symmetric shear strain.
The rotation-induced response is dominated by the intraband contribution and increases approximately linearly with the ferroaxial crystal field, vanishing in the absence of ferroaxial order.
In contrast, the symmetric-shear-induced response remains finite without ferroaxial order and contains both intraband and interband contributions.
These results demonstrate that the antisymmetric local rotation provides a distinct electronic response channel characteristic of the ferroaxial state.
}

\date{\today}
\begin{document}
\maketitle

Ferroaxial order is a ferroic state characterized by a spontaneous rotational distortion of the crystal structure, whose order parameter transforms as a time-reversal-even axial vector~\cite{Hlinka_PhysRevLett.113.165502, Hlinka_PhysRevLett.116.177602, cheong2021permutable, zeng2025photo}.
Unlike conventional ferroelectric and ferromagnetic orders, ferroaxial order can preserve both spatial inversion and time-reversal symmetries while breaking a mirror symmetry associated with the rotation axis~\cite{jin2020observation, Hayashida2020}.
The characteristic symmetry of ferroaxial order has been systematically discussed in terms of multipole representations, where the ferroaxial degree of freedom is associated with an electric-toroidal dipole~\cite{Hayami2018Multipole,Yatsushiro2021, Bhowal_PhysRevResearch.6.043141, hayami2024unified}.
Experimentally, ferroaxial order and its domain structures have been identified in a variety of materials using several experimental probes, including optical techniques such as electrogyration and electric-quadrupole second-harmonic generation~\cite{mirovitskii2014phenomenological, PhysRevLett.122.047401, PhysRevB.108.085403, shiro2026spatially}.
Representative examples include Cu$_3$Nb$_2$O$_8$~\cite{johnson2011prl_Cu3Nb2O8}, CaMn$_7$O$_{12}$~\cite{johnson2012prl_CaMn7O12}, RbFe(MoO$_4$)$_2$~\cite{jin2020observation, Hayashida2021}, NiTiO$_3$~\cite{Hayashida2020, Hayashida2021, yokota2022three, hayashida2023electric, Bhowal_PhysRevResearch.6.043141}, BaCoSiO$_4$~\cite{xu2022prb_BaCoSiO4}, Ca$_5$Ir$_3$O$_{12}$~\cite{Hasegawa_doi:10.7566/JPSJ.89.054602, hanate2021first, hayami2023cluster, hanate2023space}, K$_2$Zr(PO$_4$)$_2$~\cite{yamagishi2023ferroaxial, Bhowal_PhysRevResearch.6.043141, j6tr-ggn9}, Na$_2$Hf(BO$_3$)$_2$~\cite{nagai2023chemicalSwitching}, MnTiO$_3$~\cite{Sekine_PhysRevMaterials.8.064406}, and 1$T$-TaS$_2$~\cite{Luo_PhysRevLett.127.126401, liu2023electrical}.
More recently, characteristic electronic responses associated with ferroaxial order have also been theoretically explored through its coupling to electronic degrees of freedom~\cite{he2026universal}, such as longitudinal spin-current generation~\cite{hayami2022electric, roy2022unconventional, yamanaka2026thermodynamic}, antisymmetric thermopolarization~\cite{nasu2022antisymmetric}, nonlinear transverse magnetization~\cite{inda2023nonlinear,du2026electric}, unconventional Hall effects~\cite{hayami2023unconventional}, and nonlinear Edelstein effects~\cite{m4k3-mttk}.
Ferroaxial ordering can also reconstruct the electronic structure through characteristic orbital hybridization and spin-orbit coupling~\cite{Inda2025}.

Lattice deformation provides a versatile means of modifying electronic and magnetic properties and has been extensively studied in the context of strain engineering~\cite{Li2014,Wang2019}.
In ferroaxial systems, strain-induced transport responses have also been proposed as characteristic probes of ferroaxial order~\cite{Day_PhysRevLett.134.016401, oiwa2026piezoaxial}.
In microscopic electronic models, such lattice deformation can be incorporated through deformation-induced changes in hopping amplitudes and local orbital orientations~\cite{Ogawa2023,ogawa2025nonlinear, Ogawa2026}.
Most studies of elastic responses focus on the symmetric part of the displacement gradient, namely the conventional strain tensor.
However, the displacement gradient also contains an antisymmetric component, which describes a local rotation of the lattice~\cite{LandauElasticity, Xiao2007, gopalan2011rotation}.
Thus, lattice deformation generally involves not only symmetric strain but also a rotational degree of freedom.

The antisymmetric rotational component is expected to be particularly relevant in ferroaxial systems.
Because ferroaxial order itself is characterized by an axial rotational degree of freedom, a local lattice rotation can couple to the ferroaxial electronic state in a manner distinct from conventional symmetric strain~\cite{barantani2026resonant, Xie2026Phonon}.
Nevertheless, the electronic response to the antisymmetric part of the displacement gradient itself has been much less explored than the response to conventional symmetric strain~\cite{kirikoshi2023rotational}.
Clarifying the difference between the responses to these two components of the displacement gradient is therefore essential for understanding elastic responses unique to ferroaxial systems.

In this Letter, we theoretically investigate electronic responses to symmetric strain and antisymmetric local rotation in a ferroaxial system.
We construct a tight-binding model on a two-dimensional square lattice and incorporate lattice deformation microscopically through the modulation of electronic hopping amplitudes.
By decomposing the displacement gradient into symmetric and antisymmetric components and treating them as independent perturbations, we clarify how the two types of deformation generate distinct electronic multipole responses.
In particular, we show that the antisymmetric rotational component provides a characteristic response channel associated with the ferroaxial state that is qualitatively different from that induced by conventional symmetric strain.

To investigate the above considerations microscopically, we consider a tight-binding model on a two-dimensional square lattice.
At each lattice site, we introduce four $d$ orbitals, $(d_{x^2-y^2},d_{yz},d_{zx},d_{xy})$, and include
nearest-neighbor and next-nearest-neighbor hopping processes. 
The Hamiltonian is written as
\begin{equation}
H=H_{\mathrm{hop}}+H_{\mathrm{MF}},
\end{equation}
where $H_{\mathrm{hop}}$ and $H_{\mathrm{MF}}$ represent the hopping and ferroaxial crystalline field terms, respectively.
The hopping term is expressed in momentum space as
\begin{equation}
H_{\mathrm{hop}}
=
\sum_{\bm{k}}
\sum_{\alpha,\beta=1}^{4}
\epsilon_{\alpha\beta}(\bm{k})
c_{\bm{k}\alpha}^{\dagger}c_{\bm{k}\beta},
\end{equation}
where $c_{\bm{k}\alpha}^{\dagger}$ is the creation operator for orbital $\alpha$ with momentum $\bm{k}$, with $\alpha=1,2,3,4$ corresponding to $d_{x^2-y^2}$, $d_{yz}$, $d_{zx}$, and $d_{xy}$, respectively.
The nonzero matrix elements $\epsilon_{\alpha\beta}(\bm{k})$ are given by
\begin{align}
\epsilon_{11}
&=
\frac{1}{2}(3t_{\sigma}+t_{\delta})(c_x+c_y)
+4t_{\pi}'c_xc_y,
\\
\epsilon_{22}
&=
2t_{\pi}c_y
+2t_{\delta}c_x
+2(t_{\pi}'+t_{\delta}')c_xc_y,
\\
\epsilon_{23}
&=
-2(t_{\pi}'-t_{\delta}')s_xs_y,
\\
\epsilon_{33}
&=
2t_{\pi}c_x
+2t_{\delta}c_y
+2(t_{\pi}'+t_{\delta}')c_xc_y,
\\
\epsilon_{44}
&=
2t_{\pi}(c_x+c_y)
+(3t_{\sigma}'+t_{\delta}')c_xc_y,
\end{align}
where $\epsilon_{32}(\bm{k})=\epsilon_{23}(\bm{k})$ and all other matrix elements vanish.
For notational simplicity, we define $c_\mu=\cos k_\mu$ and $s_\mu=\sin k_\mu$ for $\mu=x,y$.

To realize a ferroaxial state characterized by broken vertical mirror symmetry, we introduce a crystal-field term that hybridizes the $d_{x^2-y^2}$ and $d_{xy}$ orbitals,
\begin{equation}
H_{\mathrm{MF}}
=
h\sum_{\bm{k}}
\left(
c_{\bm{k}1}^{\dagger}c_{\bm{k}4}
+
c_{\bm{k}4}^{\dagger}c_{\bm{k}1}
\right),
\end{equation}
Here, $h$ denotes the ferroaxial crystal field, which serves as the ferroaxial order parameter and lowers the point-group symmetry from $D_{\rm 4h}$ to $C_{\rm 4h}$.
This term hybridizes the $d_{x^2-y^2}$ and $d_{xy}$ orbitals and represents the ferroaxial degree of freedom in the present model.


We next consider the modulation of the nearest-neighbor and next-nearest-neighbor hopping integrals induced by lattice deformation.
A time-dependent lattice displacement with wave vector $\bm{q}$ and frequency $\omega$ is introduced as $\delta R_{i,\bm{q},\omega} e^{i(\bm{q}\cdot\bm{r}-\omega t)}$ for $i=x,y$.
Following Refs.~\citen{Froyen1979,Harrison1980}, we assume a phenomenological distance dependence $t(R) \propto R^{-2}$.
For the present four-orbital model, the terms not involving the $d_{x^2-y^2}$ orbital follow Ref.~\citen{Ogawa2026}, while those involving the $d_{x^2-y^2}$ orbital are derived in the present study.
We retain only the terms linear in the displacement gradient, i.e., the leading-order terms in both the displacement amplitude and the wave vector $\bm{q}$.
In the following, we focus on the symmetric shear strain $u_{xy}$ and the antisymmetric local rotation $\theta_z$, defined as
$u_{xy}=(\partial_x\delta R_y+\partial_y\delta R_x)/2
\simeq (iq_x\delta R_{y,\bm{q},\omega}
+iq_y\delta R_{x,\bm{q},\omega})/2$
and
$\theta_z=(\partial_x\delta R_y-\partial_y\delta R_x)/2
\simeq i(q_x\delta R_{y,\bm{q},\omega}-q_y\delta R_{x,\bm{q},\omega})/2$.

The resulting perturbation Hamiltonian is written as
\begin{equation}
\delta H =
\sum_{\bm{k}} \sum_{\alpha,\beta=1}^{4}
\delta\epsilon_{\alpha\beta}(\bm{k};\bm{q})
c_{\bm{k}+\frac{\bm{q}}{2},\alpha}^{\dagger}
c_{\bm{k}-\frac{\bm{q}}{2},\beta}
e^{-i\omega t}.
\end{equation}
where 
$\delta\epsilon_{\alpha\beta}(\bm{k};\bm{q})$ denotes the matrix element of the deformation-induced perturbation $\delta\mathcal{H}(\bm{k};\bm{q})$ in the orbital basis, i.e., $\delta\mathcal{H}(\bm{k};\bm{q}) = \sum_{\alpha,\beta} \delta\epsilon_{\alpha\beta}(\bm{k};\bm{q}) |\alpha\rangle\langle\beta|$.
For the $u_{xy}$ and $\theta_z$ components considered here, the nonzero matrix elements are
\begin{align}
\delta\epsilon_{11}
&=
8t_{\pi}'s_xs_y\,u_{xy},
\\
\delta\epsilon_{14}
&=
\left[
t(c_x+c_y)
-2t' c_xc_y
\right]\theta_z
+t(c_x-c_y)\,u_{xy},
\\
\delta\epsilon_{22}
&=
4 t'_{-} s_xs_y\,\theta_z
-4 t'_{+} s_xs_y\,u_{xy},
\\
\delta\epsilon_{23}
&=
-2 t_{-}(c_x-c_y)\,\theta_z
-\left[
2 t_{-}(c_x+c_y)
-4 t'_{-} c_xc_y
\right]u_{xy},
\\
\delta\epsilon_{33}
&=
-4 t'_{-} s_xs_y\,\theta_z
-4 t'_{+}s_xs_y\,u_{xy},
\\
\delta\epsilon_{44}
&=
-2(3t_{\sigma}'+t_{\delta}')s_xs_y\,u_{xy},
\end{align}
with $\delta\epsilon_{41}=\delta\epsilon_{14}$,
$\delta\epsilon_{32}=\delta\epsilon_{23}$, and all other matrix elements vanishing. 
Here, we have introduced $t^{(')} = 3t^{(')}_{\sigma}-4t^{(')}_{\pi}+t^{(')}_{\delta}$ and $t^{(')}_{\pm} = t^{(')}_{\pi} \pm t^{(')}_{\delta}$ for compactness. 
It is noted that the antisymmetric rotation $\theta_z$ does not modify the bond length to linear order.
Its contribution to the hopping modulation instead originates from the change in the relative orientation between the bond direction and the local $d$-orbital basis.
The complete expressions including the remaining strain components are given in the Supplemental Material.

We now investigate the electronic response to lattice deformation.
To characterize the electronic degrees of freedom induced by the deformation, we focus on electric quadrupole moments.
We therefore evaluate the correlation between the deformation-induced perturbation and the electric quadrupole operator within linear response theory~\cite{Kubo1957}.
The corresponding dynamical susceptibility is given by
\begin{equation}
\begin{aligned}
\chi_{\gamma}(\bm{q},\omega)
&=
\frac{1}{N}
\sum_{\bm{k}}\sum_{m,n}
\frac{
f(\varepsilon_{m\bm{k}})
-f(\varepsilon_{n\bm{k}})
}{
\omega+\varepsilon_{m\bm{k}}
-\varepsilon_{n\bm{k}}+i\delta
}
\\
&\quad\times
\langle m\bm{k}|
\mathcal{Q}_{\gamma}
|n\bm{k}\rangle
\langle n\bm{k}|
\delta\mathcal{H}(\bm{k};\bm{q})
|m\bm{k}\rangle ,
\end{aligned}
\end{equation}
where $\varepsilon_{n\bm{k}}$ and $|n\bm{k}\rangle$ denote the eigenenergy and eigenstate of the unperturbed Hamiltonian for the band $n$ and the wave vector $\bm{k}$, respectively, and $f(E)=[e^{\beta(E-\mu)}+1]^{-1}$ is the Fermi--Dirac distribution function with the chemical potential $\mu$.
Here, $\mathcal{Q}_{\gamma}$ is the one-particle representation of the electric quadrupole operator
\begin{equation}
Q_{\gamma}
=
\frac{1}{N}\sum_{\bm{k}}
\sum_{\alpha,\beta=1}^{4}
c_{\bm{k}\alpha}^{\dagger}
\mathcal{Q}^{\gamma}_{\alpha\beta}
c_{\bm{k}\beta},
\end{equation}
where $\gamma=(xx,yy, zz, yz, zx, xy)$ labels the quadrupole component. 
The quadrupole operators are defined according to Ref.~\citen{Kusunose_JPSJ.77.064710}, and their explicit matrix
representations are given in the Supplemental Material.
The matrix elements of the deformation-induced one-particle perturbation $\delta\mathcal{H}(\bm{k};\bm{q})$ are $\delta\epsilon_{\alpha\beta}(\bm{k};\bm{q})$ introduced above.
The parameter $\delta$ is a phenomenological broadening parameter describing relaxation, and $N$ denotes the number of lattice sites.

To extract the response linear in the frequency of the lattice deformation, we define
\begin{equation}
\Phi_{\gamma}(\bm{q})
=
\lim_{\omega\rightarrow0}
\frac{
\chi_{\gamma}(\bm{q},\omega)
-
\chi_{\gamma}(\bm{q},0)
}{\omega}.
\end{equation}
The quantity $\Phi_{\gamma}$ characterizes the low-frequency dynamical electric-quadrupole response induced by lattice deformation.
To distinguish the electronic responses to the symmetric and antisymmetric components of the displacement gradient, we treat $u_{xy}$ and $\theta_z$ as independent perturbations and evaluate $\Phi_{\gamma}$ separately for each case.
Throughout the calculations, we set the hopping parameters to $(t_{\sigma},t_{\pi},t_{\delta},t_{\sigma}',t_{\pi}',t_{\delta}')=(1.0,0.5,0.2,0.1,0.05,0.02)$ and the inverse temperature to $\beta=1000$.
We decompose $\Phi_{\gamma}$ into intraband and interband contributions, corresponding to transitions within the same energy band ($\varepsilon_{n\bm{k}}=\varepsilon_{m\bm{k}}$) and between different energy bands ($\varepsilon_{n\bm{k}}\neq\varepsilon_{m\bm{k}}$), respectively.

\begin{figure}[t]
    \centering
    \includegraphics[width=1.0\linewidth]{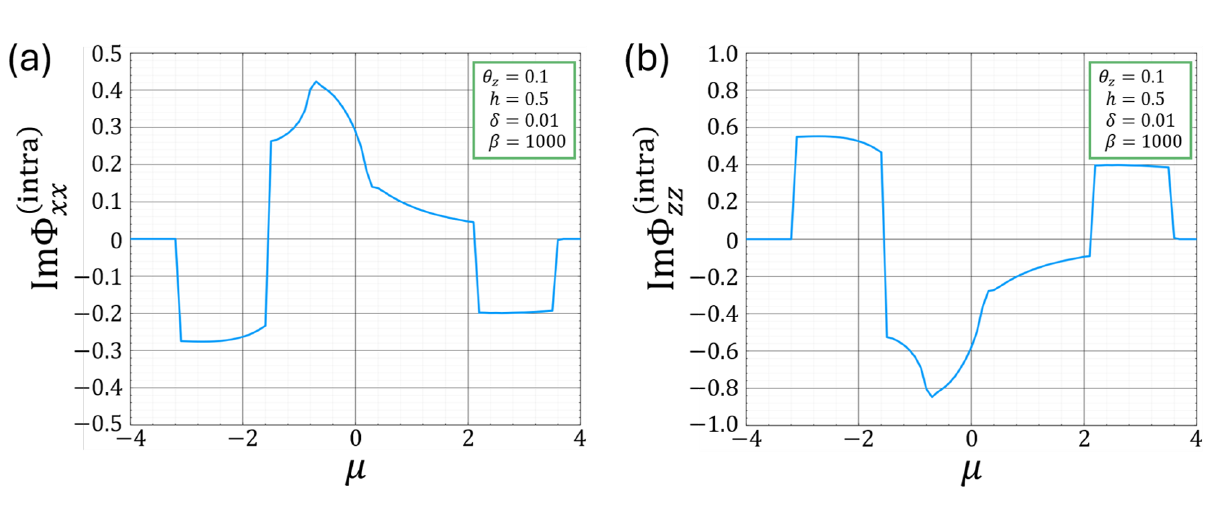}
    \caption{(Color online) 
    Chemical-potential dependence of the intraband (a) $\Phi_{xx}$ and (b) $\Phi_{zz}$ responses induced by the
    antisymmetric local rotation $\theta_z$.
    }\label{fig:mu}
\end{figure}

Before presenting the numerical results, we discuss the symmetry-allowed quadrupole responses to $u_{xy}$ and $\theta_z$.
In the absence of ferroaxial order ($h=0$), the system belongs to the $D_{\rm 4h}$ point group.
The symmetric shear strain $u_{xy}$ and the electric quadrupole $Q_{xy}$ transform according to the same $B_{2g}$ irreducible representation, and hence the response $Q_{xy}\propto u_{xy}$ is already allowed in the $D_{\rm 4h}$ phase.

The situation is qualitatively different for the antisymmetric local rotation $\theta_z$.
The ferroaxial ordering lowers the point-group symmetry from $D_{\rm 4h}$ to $C_{\rm 4h}$ by breaking the vertical mirror symmetries.
In the multipole representation, this symmetry reduction activates the $z$ components of the electric-toroidal dipole, $G_z$, and the electric-toroidal octupole, $G_z^\alpha$, which are forbidden as totally symmetric multipoles in $D_{\rm 4h}$ but become active in $C_{\rm 4h}$~\cite{Hayami2018Multipole}.
Since the local rotation $\theta_z$ has the same axial transformation property as $G_z$ and $G_z^\alpha$, the quadrupole response to $\theta_z$ is described by an axial response tensor $\Phi$.
Its symmetry-allowed components in the ferroaxial $C_{\rm 4h}$ phase include the diagonal components satisfying
$\Phi_{xx}=\Phi_{yy}$ and $\Phi_{zz} \neq \Phi_{xx}$ for the electric-toroidal dipole and $\Phi_{xx}=\Phi_{yy}$ and $\Phi_{zz}=-2\Phi_{xx}$ for the electric-toroidal octupole~\cite{Yatsushiro2021}.
These components are absent in the parent $D_{\rm 4h}$ phase.

Equivalently, this symmetry-allowed response can be understood from the viewpoint of the parent $D_{\rm 4h}$ symmetry.
Both the antisymmetric local rotation $\theta_z$ and the ferroaxial order parameter $h$ transform according to the $A_{2g}$ irreducible representation.
Their product $h\theta_z$ therefore belongs to the fully symmetric $A_{1g}$ representation, allowing diagonal electric-quadrupole components satisfying $\Phi_{xx}=\Phi_{yy}$ to be induced.
The detailed relation between the in-plane and out-of-plane components reflects the relative contributions of the
electric-toroidal dipole $G_z$ and octupole $G_z^\alpha$ in the $C_{\rm 4h}$ phase.

Since this response requires the ferroaxial order parameter, its leading contribution is expected to be odd in $h$ and approximately linear in $h$ in the weak-order regime.
This behavior is in marked contrast to the $Q_{xy}$ response to the symmetric shear strain $u_{xy}$, which is already symmetry allowed in the parent $D_{\rm 4h}$ phase and therefore remains finite even at $h=0$.
Thus, the contrasting $h$ dependences of the responses to $\theta_z$ and $u_{xy}$ provide a direct manifestation of the symmetry reduction from $D_{\rm 4h}$ to $C_{\rm 4h}$ associated with ferroaxial ordering.

We first examine the response to the antisymmetric local rotation $\theta_z$.
Figure~\ref{fig:mu} shows the chemical-potential dependence of the intraband contributions to $\mathrm{Im}[\Phi_{xx}]$ and $\mathrm{Im}[\Phi_{zz}]$, which describe the induction of the diagonal electric quadrupoles $Q_{xx}$ and $Q_{zz}$, respectively.
Finite responses appear over a broad range of the chemical potential.
We also obtain $\Phi_{xx}=\Phi_{yy}$, whereas the responses of the other quadrupole components are negligibly small.
Furthermore, the calculated responses satisfy $\Phi_{zz}=-2\Phi_{xx}$.

The relation $\Phi_{xx}=\Phi_{yy}$ is consistent with the tetragonal symmetry of the ferroaxial $C_{\rm 4h}$ phase, while the additional relation $\Phi_{zz}=-2\Phi_{xx}$ corresponds to the tensor structure associated with the electric-toroidal octupole $G_z^\alpha$ discussed above~\cite{Yatsushiro2021}.
Thus, among the symmetry-allowed $G_z$- and $G_z^\alpha$-type contributions, the present model selectively realizes the $G_z^\alpha$-type diagonal response.
This is consistent with the previous study based on a five-$d$-orbital model, in which the expectation value of the electric-toroidal dipole $G_z$ was found to vanish despite the presence of ferroaxial order, while higher-rank electric-toroidal multipoles remain active~\cite{hayami2022electric}.
Thus, symmetry alone allows both $G_z$- and $G_z^\alpha$-type contributions, whereas their relative weights are determined by the microscopic orbital structure.

\begin{figure}[t]
    \centering
    \includegraphics[width=1.0\linewidth]{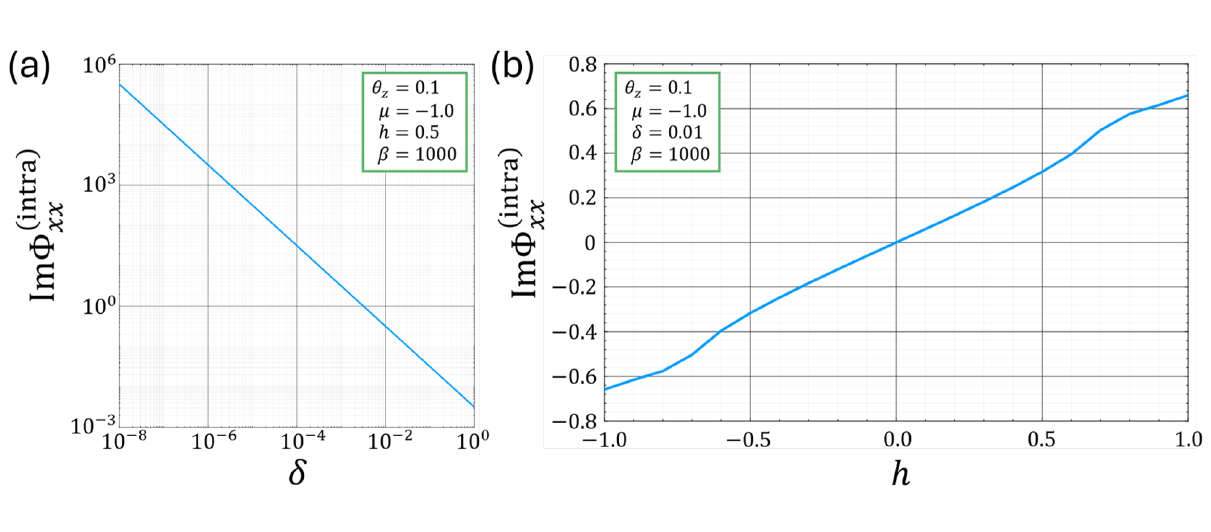}
    \caption{(Color online) 
    (a) Broadening-parameter $\delta$ dependence of the intraband response $\Phi_{xx}$ on a log-log scale.
    (b) Ferroaxial crystal-field $h$ dependence of the intraband response $\Phi_{xx}$.
    }\label{fig:h}
\end{figure}

The interband contribution to the rotation-induced response is found to be negligibly small over the parameter range considered here, showing that the response is dominated by intraband processes. 
To further clarify its microscopic character, we examine the dependence of $\Phi_{xx}$ on the phenomenological broadening $\delta$ and the ferroaxial crystal field $h$, as shown in Fig.~\ref{fig:h}.

Figure~\ref{fig:h}(a) shows that the magnitude of $\Phi_{xx}$ scales approximately as $\delta^{-1}$.
Such an inverse-broadening dependence is characteristic of an intraband contribution and indicates that the response is governed predominantly by electronic states near the Fermi level.
Since $\delta^{-1}$ plays the role of a relaxation time within the present treatment, the result also shows that the
rotation-induced quadrupole has a relaxation-time-dependent, transport-like character, rather than being dominated by virtual interband transitions.

Figure~\ref{fig:h}(b) shows the dependence of $\Phi_{xx}$ on the ferroaxial crystal field $h$.
In the weak-$h$ regime, $\Phi_{xx}$ increases approximately linearly with $h$ and approaches zero as $h\to0$.
This behavior is consistent with the symmetry analysis above: the diagonal quadrupole response to $\theta_z$ arises from the combined presence of the ferroaxial order and the antisymmetric local rotation, and its leading contribution is therefore proportional to $h\theta_z$.
The strong correlation between the response amplitude and $h$ demonstrates that the rotation-induced quadrupole directly reflects the ferroaxial electronic state.


\begin{figure}[t]
    \centering
    \includegraphics[width=1.0\linewidth]{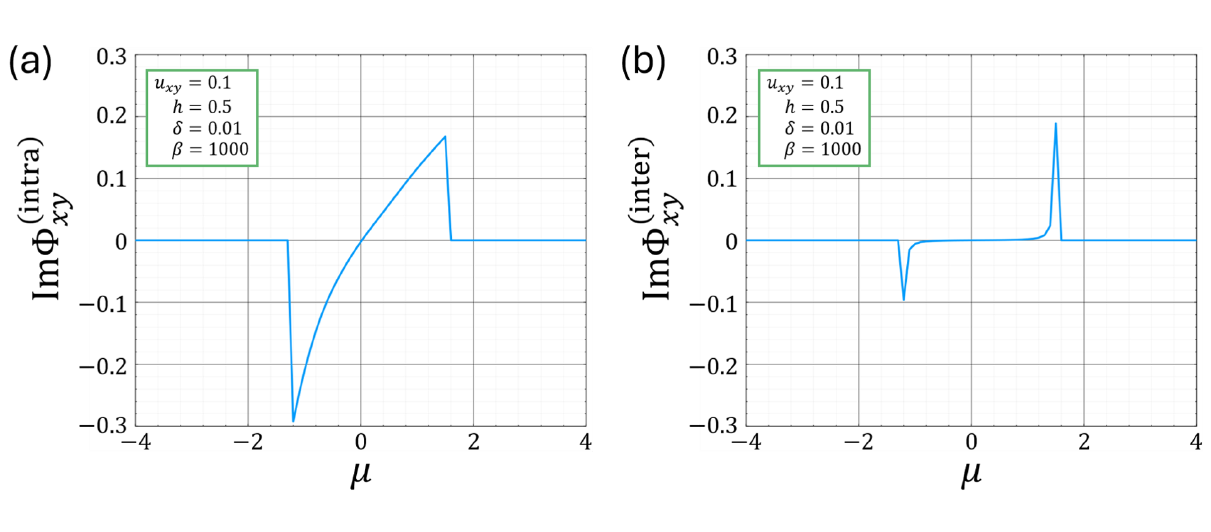}
    \caption{(Color online) 
    Chemical-potential dependence of the (a) intraband and (b) interband contributions to $\mathrm{Im}[\Phi_{xy}]$ induced by the symmetric shear strain $u_{xy}$.
    }\label{fig:mu_uxy}
\end{figure}

We next examine the response to the symmetric shear strain $u_{xy}$. 
Figure~\ref{fig:mu_uxy} shows the chemical-potential dependence of the intraband and interband contributions to $\mathrm{Im}[\Phi_{xy}]$, corresponding to the induced electric quadrupole $Q_{xy}$.
In contrast to the response to $\theta_z$, only the off-diagonal quadrupole $Q_{xy}$ shows an appreciable response, while the other quadrupole components remain negligibly small.
Both the intraband and interband contributions are finite and exhibit distinct chemical-potential dependences.

This behavior is consistent with the symmetry argument given above. 
Since $u_{xy}$ and $Q_{xy}$ belong to the same $B_{2g}$ irreducible representation in the parent $D_{\rm 4h}$ phase, their coupling does not require ferroaxial symmetry breaking.
Accordingly, the $Q_{xy}$ response remains finite even in the $h\to0$ limit and shows only a weak dependence on the ferroaxial crystal field.
The finite interband contribution further distinguishes this response from the $\theta_z$-induced diagonal response, which is almost entirely of intraband origin.

The contrasting behaviors for $u_{xy}$ and $\theta_z$ therefore reflect their different symmetry origins.
The symmetric shear strain directly couples to the electric quadrupole $Q_{xy}$ already in the parent $D_{\rm 4h}$ phase, whereas the antisymmetric local rotation activates the diagonal quadrupole response only after the ferroaxial symmetry reduction to $C_{\rm 4h}$.
The nearly linear dependence on $h$ and the intraband-dominated nature of the latter response provide characteristic signatures that distinguish the local-rotation response from the conventional symmetric-strain response.


In summary, we have theoretically investigated the electronic response to lattice deformation in a ferroaxial system by explicitly distinguishing the symmetric and antisymmetric components of the
displacement gradient.
Using a four-orbital tight-binding model on a two-dimensional square lattice, we incorporated lattice deformation microscopically through the modulation of electronic hopping amplitudes and evaluated the induced electric quadrupoles within linear response theory.
We found qualitatively distinct responses to the antisymmetric local rotation and the symmetric shear strain. 
The antisymmetric local rotation induces diagonal electric quadrupoles only in the ferroaxial $C_{\rm 4h}$ phase.
The resulting response satisfies $\Phi_{xx}=\Phi_{yy}$ and $\Phi_{zz}=-2\Phi_{xx}$, corresponding to the tensor structure associated with the electric-toroidal octupole $G_z^\alpha$.
Its approximately linear dependence on the ferroaxial crystal field demonstrates that this response directly reflects the ferroaxial order. 
Moreover, the rotation-induced response is dominated by the intraband contribution and exhibits an inverse dependence on the broadening parameter.
In contrast, the symmetric shear strain directly induces the off-diagonal electric quadrupole $Q_{xy}$, as already allowed by the parent $D_{\rm 4h}$ symmetry.
The corresponding response remains finite even in the absence of ferroaxial order, and both intraband and interband processes contribute to the response.
These contrasting behaviors demonstrate that the antisymmetric local rotation and symmetric shear strain, although originating from the same displacement gradient, couple to electronic degrees of freedom through distinct symmetry channels.

This research was supported by JSPS KAKENHI Grant Numbers JP22H00101, JP23H04869, and JP26K22277, and by JST CREST (JPMJCR23O4) and JST FOREST (JPMJFR2366).

\bibliography{Ferroaxial_ver2.bib}
\bibliographystyle{jpsj}

\end{document}